\documentclass[10pt,conference,a4paper]{IEEEtran}
\usepackage{graphicx}
\begin{document}

\title{Video Surveillance Robot Powered by Raspberry Pi}
%\titlenote{Produces the permission block, and
%  copyright information}
%\subtitle{Extended Abstract}
%\subtitlenote{The full version of the author's guide is available as
%  \texttt{acmart.pdf} document}

%\author{Jonay Su\'arez-Armas}
%\authornote{The corresponding author}
%\orcid{0000-0003-3022-0357}
%\email{jsuarear@ull.edu.es}
%\author{Pino Caballero-Gil}
%\email{pcaballe@ull.edu.es}
%\author{C\'andido Caballero-Gil}
%\email{ccabgil@ull.edu.es}
%\affiliation{%
%  \institution{University of La Laguna}
%  \department{Department of Computer Engineering and Systems}
%  \city{La Laguna}
%  \state{Canary Islands}
%  \postcode{38205}
%  \country{Spain}
%}
\author{Jonay Su\'arez-Armas, Pino Caballero-Gil, C\'andido Caballero-Gil\\
University of La Laguna. Canary Islands. Spain\\ \{jsuarear, pcaballe, ccabgil\}@ull.edu.es}

\maketitle
\begin{abstract}
Video surveillance systems are increasingly used in different fields, from the domestic  to the commercial environment. Current systems are being improved and complemented with new elements and functionalities. This paper proposes the design of a video surveillance robot based on Raspberry Pi with the abilities to perform tasks of motion detection, send video on real time, fire detection and also, the possibility of control it remotely from the Internet. In order to check the information received from the robot, as well as the video sent, a client application has been developed to any device with an Internet connection. In addition to this, in order to protect the information obtained by the robot, a secure system is proposed, which uses different security mechanisms to achieve this goal.
\end{abstract}

\section{Introduction}\label{sec.introduction}
\begin{figure}[ht]
	\centering
	\includegraphics[width=0.35\textwidth]{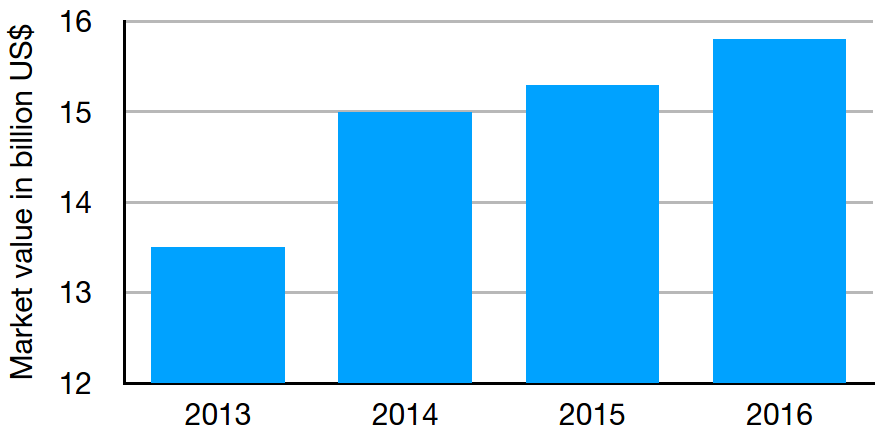}
	\caption{Investment in video surveillance systems}
	\label{graphvs}
\end{figure}

Nowadays, the interest in physical security at homes, commercial establishments and big enclosures has risen the number of video surveillance systems that every year are installed \cite{stad}. The graph presented in Fig.~\ref{graphvs} shows the increase in money invested in video surveillance systems last years all over the world. This is due to the fact that the number of thefts has also increased, and people need to protect their property and even their lives.

Currently installed video surveillance systems are equipped with fixed cameras that allow to check a certain zone, or, in some cases cameras that can be focused remotely to different zones, depending on the capacities of them. This means that the cameras can not be moved to explore an area in more detail, but it is possible to enlarge the images in some models. By combining video surveillance and robotics, a solution that allows moving cameras along enclosures can be achieved, so, at this way, more information can be obtained from the protected areas. In addition, the wireless technologies allow to move the cameras to any place and to obtain the information of them.

In this paper, the design of a video surveillance robot based on Raspberry Pi that is able of performing tasks of motion detection, sending live video and fire detection is proposed. This project starts from an earlier version \cite{robocam} in which a smartphone is used to capture images, it is connected via Bluetooth with a Lego Mindstorms robot that holds such smartphone. In addition to this, there is a server that allows to send it the information and control the robot remotely. This version improves the previous one by using a Raspberry Pi to replace the expensive Lego Mindstorms robot and the smartphone.

The next section shows some related works. Section \ref{sec.design} explains the design of the proposed prototype, while in Section \ref{sec.functionalities} the details of the functionalities that the robot will be able to perform are presented. The communications between the different elements of the system are explained in section \ref{sec.communications}. Finally, the paper is closed with a brief conclusion.

\begin{figure*}[ht]
	\centering
	\includegraphics[width=0.63\textwidth]{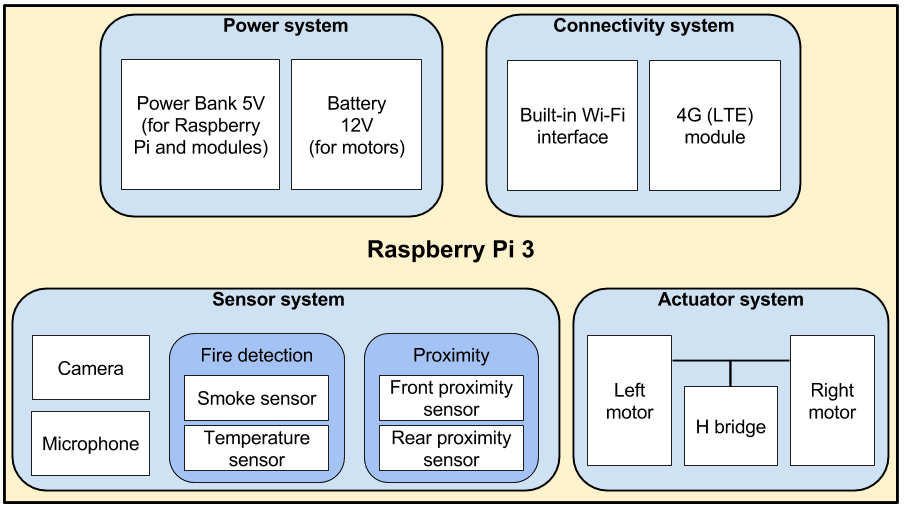}
	\caption{Robot scheme}
	\label{robotscheme}
\end{figure*}

\section{Related works}\label{sec.relatedworks}
There are different projects that works some of the different parts here proposed.

A proposal for motion detection is the technique called Background Subtraction \cite{backgroundsubtraction} in which the idea is to take a reference image called background to compare the frames of the video with its. For this algorithm, a threshold is established, and in the comparison of the frames with the background it is necessary to overcome that threshold to determine that there has been motion in front of the camera. There are also advanced algorithms of motion detection that allow to discard small changes in the images \cite{motdetalg}. With this, it is possible to obviate changes of light or irrelevant changes and so as not to give a false alarm. Some works also use a Raspberry Pi together with OpenCV to perform motion detection tasks \cite{rpiopencvmotdet}.

The constant improvements in the networks have caused that audio and video streaming is increasingly used, especially in social networks. For that reason, many authors work on improving the protocols to allow better audio and video quality while delay time is reduced. In \cite{streaming}, different streaming systems and platforms are analyzed.

The incorporation of sensors to Raspberry Pi provides multiple use possibilities. Among them it is possible to find the detection of fires thanks to the use of smoke \cite{smokedetection} or temperature sensors, even there are works that use Raspberry Pi together with Arduino boards that incorporates cameras that allow the user checks if really there is a fire at controlled area, and that allow to send notifications via SMS \cite{fireraspbarduino}. Another use case consists of using proximity sensors to detect an object before colliding with it. On this there are projects that can be applied to vehicles as parking sensors that also incorporate a camera to obtain images on a screen that help the driver \cite{parkingsensor}.

The use of Raspberry Pi and Arduino microcontrollers in the field of Internet of Things (IoT) is constantly rising, so every day there are more works that use devices of these types. In \cite{arduinoelderlyrobot}, the design of an Arduino-based robot to monitor elderly people is proposed, which is able to follow people thanks to pre-saved maps and ultrasound. In \cite{raspberryobstaclesrobot}, the authors propose a robot based on Raspberry Pi that use a camera to detect obstacles thanks to an algorithm and it moves in a direction in which there is no obstacle. There are also other proposals that combine both devices to have more possibilities of use, such as in \cite{arduinorasprobot}, where a Raspberry Pi is used with an Arduino to build a robot with vehicle form that has different type of sensors.

The functionalities of these last works (according to the reference number) are related with the proposal of this paper in the Table~\ref{relatedfunc}.

\begin{table}
	\centering
	\begin{tabular}{|c|c|c|c|c|c|c|}
		\hline
		\textbf{Function / Reference} 	& 			[5]        	& 			[2]		& 		[4]			& 			[10]		& 			[8]		& 				This			\\ \hline
		Obstacles avoiding					&  					&						&			X			&		X				&			X			&				X				\\ \hline
		Fire detection							&  		X			&			X			&						&						&						&				X				\\ \hline
		Motion detection						&  					&						&						&						&						&				X				\\ \hline
		Notifications								&  					&			X			&						&						&						&				X				\\ \hline
		Streaming									&  					&						&						&						&						&				X				\\ \hline
		Remote control							&  					&						&			X			&						&						&				X				\\ \hline
		Backup										&  					&						&						&						&						&				X				\\ \hline
		Indoor location							&  					&						&			X			&						&						&								\\ \hline
		Images sending							&  					&			X			&						&						&						&				X				\\ \hline
	\end{tabular}
	\newline
	\caption{Relation of the functionalities.}
	\label{relatedfunc}
\end{table}

\section{Design}\label{sec.design}
The designed robot is based on Raspberry Pi and is composed of 4 systems (see Fig.~\ref{robotscheme}): power system, connectivity system, sensor system and actuator system.

%\begin{figure*}[ht]
%	\centering
%	\includegraphics[width=0.8\textwidth]{robot5}
%	\caption{Robot scheme}
%	\label{robotscheme}
%\end{figure*}

The power system of the robot consists of two elements. The first of them is a Power Bank that provides 5V DC, and is used to power the Raspberry Pi. The second is a 12V DC battery to power the motors that allow to move the robot.

In order to obtain information from the robot, it is necessary that it is connected to the Internet, and the connectivity system is in charge of providing a way to do it. This system allows the connection in two different ways. On the one hand, it is possible to connect the robot to the Internet through a Wi-Fi connection using the built-in interface of the Raspberry Pi. On the other hand, a 4G module (LTE), to have an Internet connection in places where a Wi-Fi network is not available thanks to the mobile telephone network, has been incorporated.

The main function of the designed prototype is to collect information in different ways, so it is necessary to incorporate a sensor system, which has different types of sensors. Being a video surveillance robot, the main sensor that incorporates is a camera that allows to capture images and video, as well as a microphone to capture the audio of the environment. This system is also composed of two other subsystems. Firstly, the fire detection system with a smoke sensor and a temperature sensor, secondly, a proximity subsystem that is composed of two proximity sensors (front and rear) to avoid the collision of the robot with an obstacle.

The main functions of the robot are linked to some type of movement, so it is essential to include an actuator system in the design. This system is where the motors to move the robot are included, and for that reason an H-bridge is incorporated to make the connection of the motors with the Raspberry Pi. In addition, the motors are connected to the 12V DC battery previously named in the power system.

\section{Functionalities}\label{sec.functionalities}
The prototype has been designed to perform 4 functionalities, although it is possible to add new ones in the future. These features are: motion detection, audio and video streaming, fire detection, and remote control.

Motion detection is an essential functionality in video surveillance systems, because thanks to it, it is possible to send alerts when an unwanted presence is detected in the monitored area. In addition, this also allows to save the necessary video fragments, that is to say, in which an intrusion has occurred, discarding the rest of the video. The videos captured when motion is detected are stored and can be viewed at any other time. To carry out this functionality, the prototype uses OpenCV, which can be executed in Raspberry Pi and has a great number of functionalities and alternatives of use, and one of them is motion detection. This mode allows to send notifications to the users when motion detection occurs, so that they can see what is happening. That notification includes the video captured at the moment of the detection and a link to view the images in real time via streaming. In this way it is possible to check if the detection is real or is a false alarm.

Moreover, the camera is used to capture video in real time and send the images via streaming, so that they can be viewed in real time from anywhere, simply by using a device with an Internet connection and a web browser. Thanks to a microphone, the robot also sends the audio of the environment. The sending of audio and video to the streaming server is over the RTSP \cite{rtsp} protocol, and the reception from the server to the devices where the images will be view is over the RTMP protocol.

Apart from the functionalities described above, the robot also allows being remotely controlled over the Internet. To do this, it is possible to connect it to a Wi-Fi network in environments that have that type of connection or also to a mobile network through 4G in open environments where there is not a connection of the previous type. From a web application it is possible to obtain information about the state of the robot and its sensors, as well as receive in real time the images captured by the camera that it incorporates and also send a series of signals that allow to move it through the controlled environment. In the client application, it is possible to view the state of all the robots connected to the server and the video captured by their cameras at any time. Each robot has a window in the web application with that information and with a buttons to send commands to it.

To prevent data loss when the robot loses the Internet connection there is a backup system, which saves the information until the connection is recovered. This implies that the videos captured without connection can be visualized later when these are uploaded.

\begin{figure}[h]
	\centering
	\includegraphics[width=0.39\textwidth]{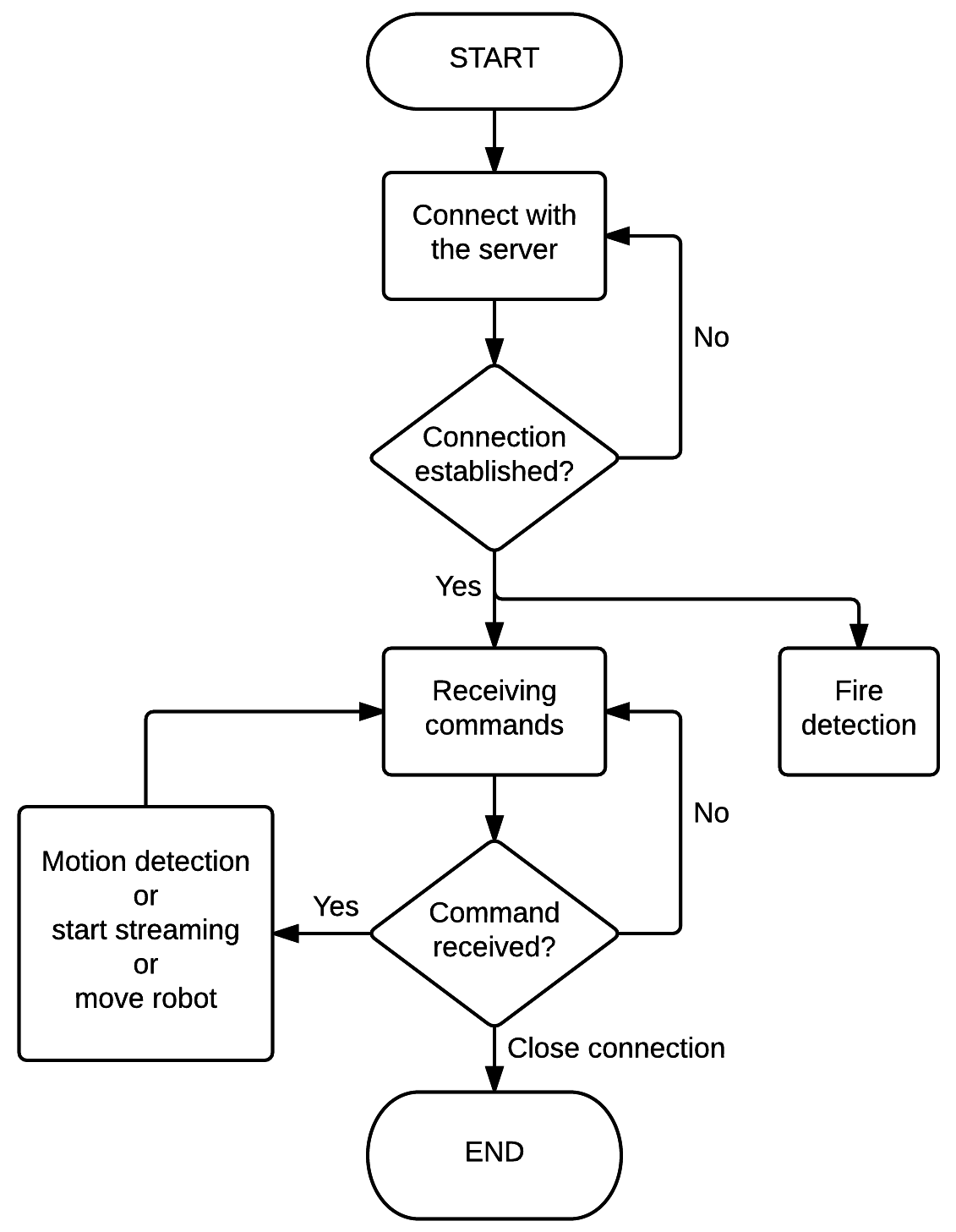}
	\caption{Operation flowchart}
	\label{operation}
\end{figure}

The operation mode of the designed prototype is detailed in the flowchart of Fig.~\ref{operation}. It is possible to observe that since the robot is turned on, it establishes a connection with the server, so that a communication can always be established with it. At that moment, it is waiting to receive any kind of order and it also begins the detection of fires. The different commands that the robot can receive at that moment are: start motion detection, start streaming sending, and move in a specific direction. From the streaming and motion detection modes it is also possible to receive a moving command or to stop the current mode.

\section{Communications}\label{sec.communications}
The designed prototype is complemented by other elements that make up a video surveillance system that can be controlled through the Internet. These elements are composed by a server that hosts a web application that allows the connection with the robot through is possible to obtain the information captured by the robot in real time and send orders of movement. The server also has a streaming server installed which allows to receive the video sent by the robot and store and distribute it. In addition to this, there is a device on which the client application is run. This application allows the users to take control of the robot.

To establish a communication between the robot and the server, the first sends a web request to the server asking to open a communication channel between them. At that moment the server opens a communication socket between them, so that both are communicated in order to receive and send the corresponding data. This occurs when starts the operating system of the Raspberry Pi, and for this it is necessary to configure the network to be used and the IP address of the server previously.

The communications between the server and the client devices allow to receive in the last the information sent by the robot to the server, as well as to receive the audio and the video that is sent to the streaming server.

\begin{figure}[h]
	\centering
	\includegraphics[width=0.43\textwidth]{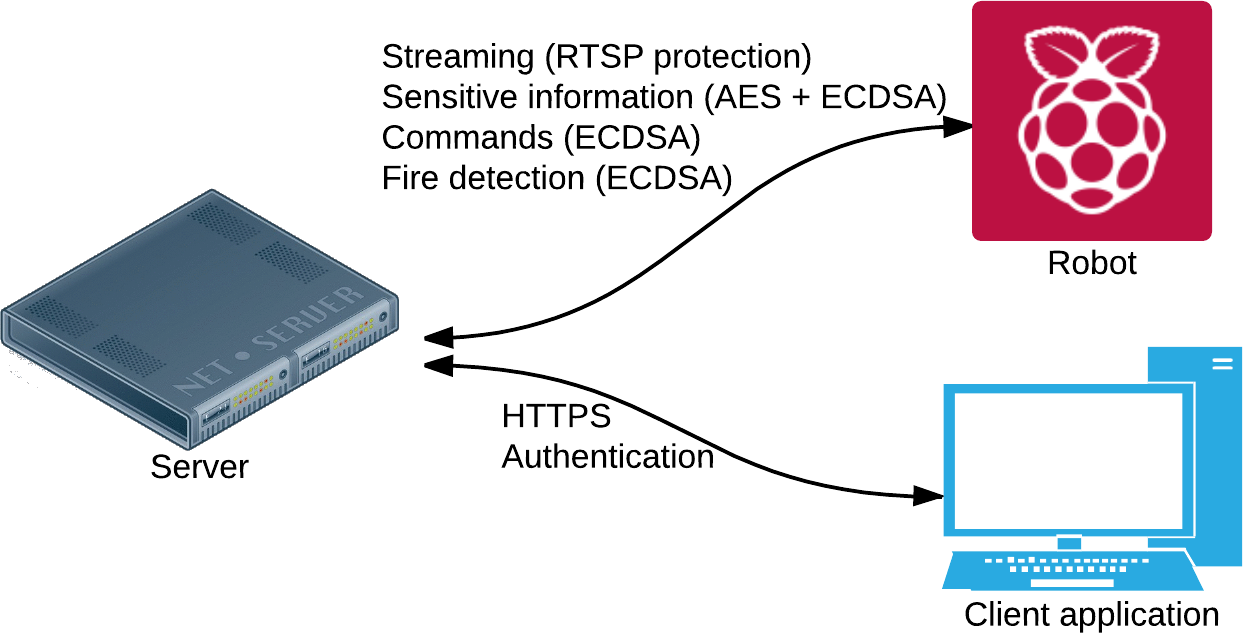}
	\caption{Security scheme}
	\label{security}
\end{figure}

%\subsection{Security}\label{subsec.security}
In order to protect the communications that occur between the elements of the system, different security mechanisms are used depending on the type of information exchanged. Fig.~\ref{security} shows the communication scheme and the security mechanisms that are applied in each part of the communications. Here the server acts as an intermediary between the robot and the client devices.

The information sent between the server and the client devices is protected by HTTPS, and the users also must be registered in the system and authenticated to access it. There is also the information that is exchanged between the robot and the server. First is the streaming video transfer, which is protected with the RTSP protocol mechanisms. The orders that are sent to the robot and the information about fire detection that the latter sends are digitally signed, and other types of information that are exchanged between them and that can be sensitive are encrypted as well as signed.

The algorithm used to encrypt the information is Advanced Encryption Standard (AES) with 256-bit key size and in Cipher Block Chaining mode. To sign and verify the data Elliptic Curve Digital Signature Algorithm (ECDSA) is used. In addition, it is necessary to make an encryption key agreement securely over the Internet, so Elliptic Curve Diffie-Hellman (ECDH) algorithm is used for that.

\section{Conclusion}\label{sec.conclusion}
Physical security of houses is something that worries people. For this reason, a video surveillance robot with cheap components and great functionality would be a great resourse. In this paper we propose such a surveillance robot which core is a mini computer Raspberry Pi. 

The robot designed has the typical video surveillance tasks and other functions such as detecting fire, gas, noises, etc. Besides, it is possible to add the functionalities and sensors necessary to control more different situations.

It can be controlled remotely and also is possible to build it in the way that best suit for the place on which it will work.

In addition to this, it can be used to view live images of an area with difficult access or in which there has been some kind of disaster or dangerous situation for a person.

The future work is to apply techniques of fuzzy logic to the motion detection, so, at this way, to discard irrelevant changes in the images that can be due to strong changes of light, to some insect in front of the camera or even the movement of some curtain due to the wind.

Taking advantage of motion detection and proximity sensors, it can also calculate how far was a movement detected. At this way, the robot could approach autonomously to this area. Whit this, better images of the affected area can be obtained.

Moreover, it is intended to perform a prototype of the robot and then make a comparative study that allows identifying the weak points to improve it later and also verify the strengths that it has.

From other work we have done which also includes streaming video, it is expected that this will be one of the points to improve, so in the future the current protocols RTSP and RTMP used for
streaming will be replaced by the WebRTC protocol \cite{webrtc}, which improve greatly the video reception delay times, becoming almost imperceptible \cite{streamingcomparative}.

\section*{Acknowledgments}
Research supported by the Spanish Ministry of Economy and Competitiveness, the European FEDER Fund, and the CajaCanarias Foundation, under Projects TEC2014-54110-R, RTC-2014-1648-8, MTM2015-69138-REDT and DIG02-INSITU.

\bibliographystyle{ACM-Reference-Format}
\bibliography{IML2017}

\end{document}